\documentclass[useAMS,usenatbib,referee]{mn2e}
\usepackage{graphicx}
\usepackage{graphics}
\usepackage{aas_macros}
\usepackage{multicol}
\usepackage{subfig}
\usepackage{etex}
\usepackage{setspace}
\usepackage{float}

\title[Phase Resolved Spectroscopy of pulsar 4U~1626--67]{Pulse phase dependence of emission lines in the X-ray pulsar 4U~1626--67}
\author[A. Beri et al.]
{Aru~Beri,$^1$
 Biswajit~Paul,$^2$ Gulab~C.~Dewangan,$^3$ \\
$^1$, Department of Physics, Indian Institute of Technology Ropar, Nangal Road, Rupnagar, Punjab, 140001 India\\
$^2$, Raman Research Institute, Sadashivnagar, C. V. Raman Avenue, Bangalore-560 080, India.\\
$^3$, Inter University Centre for Astronomy and Astrophysics, Post bag 4, Ganeshkhind, Pune, India.\\}

\begin{document}

\pagerange{\pageref{firstpage}--\pageref{lastpage}} 
\maketitle
\label{firstpage}

\begin{abstract}
We present results from a pulse phase resolved spectroscopy of the complex emission lines around 1~keV in the unique 
accretion powered X-ray pulsar 4U~1626--67 using observation made with the \emph{XMM-Newton} in 2003. In this source, the red 
and blue shifted emission lines and the line widths measured earlier with \emph{Chandra} suggest their accretion disk origin.
Another possible signature of lines produced in accretion disk can be a modulation of the line strength with pulse phase.
We found the line fluxes to have pulse phase dependence, making 4U~1626--67 only the second pulsar after Her~X--1
to show such variability. 
The O~VII line at 0.568~keV from 4U~1626--67 varied by a factor of $\sim$4, stronger than the continuum variability, that support their accretion disk 
origin. The line flux variability may appear due to variable illumination of the accretion disk by the pulsar or more likely,
 a warp like structure in the accretion disk. We also discuss some further possible diagnostics of the accretion disk in
 4U 1626-67 with pulse phase resolved emission line spectroscopy.

\end{abstract}

\begin{keywords}
X-ray: Neutron Stars - accretion, accretion disk - pulsars, individual: 4U~1626--67
\end{keywords}

\section{INTRODUCTION}

4U~1626--67, discovered with the \emph{Uhuru} satellite \citep{Giacconi72} is an accreting X-ray pulsar
with a pulse period of 7.7 seconds \citep{Rappaport77}. It is one of the very few
X-ray pulsars in low mass X-ray binary (LMXB) systems. The magnetic field strength
of the neutron star is estimated to be of B $\approx 4 \times$ $10^{12}$ Gauss \citep{Coburn02, Orlandini98, Iwakiri12} 
from the cyclotron line features in its X-ray spectrum at around 37~keV. 
This system with small mass function of $f < 1.3\times 10^{-6}M\odot $ has a neutron star
that accretes from an extremely low mass companion ($\le{0.1M}$$\odot$) %{for i=18$^\circ$) } 
\citep{Middleditch81,Levine88,Chakrabarty98}.
\citet{Jain08} reported an upper limit of 10 lt-ms for pulse arrival delay using \emph{RXTE} data.
The optical light curve shows reprocessed X-ray pulses at 130.38 mHz with multiple sidebands.
The sidebands in the power spectrum revealed an orbital period of 42 minute \citep{Middleditch81} that was confirmed by 
\citet{Chakrabarty98}. The short orbital period suggests that the binary system contains a hydrogen depleted companion 
\citep{Paczynski81}. The optical counterpart has strong UV excess and high optical pulse fraction \citep{McClintock77,
 McClintock80}. 4U~1626--67 underwent two torque reversals since its discovery. It was initially observed in a  spin-up state
 \citep{Levine88,Nagase89}, this trend reversed in 1990 and the neutron star began to spin down \citep{Wilson93, Bildsten94,
 Bildsten97, Prince94,Chakrabarty97}. After the steady spin-down phase of about 18 years, the neutron star again underwent
a torque reversal and made a transition to spin-up state in 2008 \citep{Jain10,Camero10}. The pulse profiles of this 
source show a strong correlation with its torque reversals. The pulse profiles during two spin-up eras were quite similar 
but they were different in the spin-down era \citep{Beri14}.
\\

The spectrum of 4U~1626--67 has been studied many times using data from different observatories.
 Observations with \emph{HEAO} 1 \citep{Pravdo79}, \emph{Tenma} \citep{Kii86}, \emph{ASCA} \citep{Angelini95}, 
\emph{Beppo-SAX} \citep{Orlandini98}, \emph{Chandra}, \emph{XMM-Newton} \citep{Krauss07}, \emph{RXTE} \citep{Jain10}
 and \emph{Suzaku} \citep{Camero12, Iwakiri12}  showed two component spectrum consisting of a blackbody
 and a power-law. These continuum spectral parameters showed variation during the eras of torque reversals.
 The blackbody temperature changed from 0.6 keV to 0.3 keV during first transition from spin-up to spin-down phase 
\citep{Angelini95, Vaughan97, Owens97}, followed by transition to value of $\sim$ (0.5-0.6)~keV during second phase
 of spin-up \citep{Jain10, Camero12}. During the first spin-up phase the value of powerlaw photon index was 1.5 
\citep{Vaughan97}. The spectrum became relatively harder during the spin-down phase with the powerlaw photon index
 of 0.4-0.6 \citep{Angelini95, Vaughan97, Owens97, Orlandini98, Yi99} and after the second torque reversal,
 a photon index of 0.8-1.0 was measured \citep{Jain10, Camero12}.

The presence of unusually bright Neon~(Ne) and Oxygen~(O) lines were reported from many spectroscopic observations 
\citep{Angelini95,Owens97,Schulz01,Krauss07}. The equivalent width and intensity of these emission lines were found to be variable \citep{Camero12}.
Observations made with \emph{Chandra} revealed a double peaked nature of these line features, indicating their formation in the 
accretion disc \citep{Schulz01}. The presence of these emission lines is quite interesting as these are not observed in any other
 X-ray pulsar. Oxygen emission lines are observed in both UV and optical spectra. However, optical spectrum of this source 
lacks in Neon emission features \citep{Homer02,Werner06}. 
%\cite{Orlandini98} reported an absorption feature at 37~keV using \emph{Beppo-SAX} data
% which was interpreted as Cyclotron Resonance Scattering Feature (CRSF).  This feature was later studied in detail
% using \emph{Suzaku} observations \citep{Camero12,Iwakiri12}.\\

There was a gradual decrease in the X-ray flux in the 0.3-10.0~keV band during 1977-2005 \citep{Krauss07}, followed 
by an increase in the flux during the second torque reversal \citep{Camero10,Jain10}. Observations made during spin-down
period indicated a variations in flux of low energy emission lines within the same phase of torque reversal.
The line intensities measured using \emph{ASCA} data \citep{Angelini95} were higher compared to that
measured using \emph{BeppoSAX} \citep{Owens97}, \emph{Chandra} \citep{Schulz01, Krauss07}, \emph{XMM-Newton} \citep{Krauss07}
and \emph{Suzaku} \citep{Camero12}. The fluxes of these emission lines increased during
transition from spin-down to spin-up \citep{Camero12}. \\

Pulse-Phase resolved spectroscopy is an important tool to probe into the geometry of the emission region.
The broadband X-ray continuum of the source shows a phase dependence  \citep{Pravdo79, Kii86}.
A phase dependence of 37 keV Cyclotron Resonance Scattering Feature (CRSF) was reported using \emph{Suzaku} data by \citet{Iwakiri12}, 
where they also observed it to be as an emission feature at certain phases.\\

We present here a pulse phase resolved spectral analysis of 4U~1626-67 using \emph{XMM-Newton} EPIC-pn observation made in 2003 to
show the behaviour of the  O~VII~(0.568~keV), O~VIII~(0.653~keV), Ne~IX~(0.915~keV), Ne~X~(1.021~keV)
lines in different spin phases. Using \emph{ASCA} data, \citet{Angelini95} reported that Ne Ly$\alpha$ has no spin-phase dependence.
Here we present more detailed analysis of the  emission line fluxes in narrow phase bins using \emph{XMM-Newton} observation
which has higher sensitivity than \emph{ASCA}. The paper is organised as follows: 
The details of the observation and data reduction procedures are mentioned in section 2, section-3
includes  timing analysis of the source, followed by the details of the spectroscopy. Finally, the implications of our results are
 discussed in section 5.

\section{XMM-Newton OBSERVATIONS AND DATA REDUCTION}

\emph{XMM-Newton} operates in the energy range of 0.1-15 keV. The instruments on board \emph{XMM-Newton} are European Photon Imaging
 Camera (EPIC), Reflection Grating Spectrometer (RGS) and Optical Monitor (OM). EPIC consists of two MOS \citep{Turner01} and one
 PN \citep{Struder01} CCD arrays having moderate spectral resolution and a time resolution in the range of micro-seconds to 2.6 seconds depending on the instrument and the mode of observation. 
High-resolution spectral information (E/dE $\approx$ 300) is provided by the Reflection Grating Spectrometer (RGS) \citep{Herder01}.
 Simultaneous optical/UV observations are made with Optical Monitor (OM).\\

We have used the longest observation (84~ks) of 4U~1626--67 made with \emph{XMM-Newton} on 20 August 2003 (ObsID-0152620101). The EPIC-pn
data of this observation were collected in the small window mode and using the medium filter.\\ 
 
\emph{HEASOFT~6.12} and \emph{SAS-12.0.1} were used for the reduction and extraction. For the extraction of the 
cleaned EPIC-pn events, standard filters were applied. The observation was checked for the particle background
 using the \emph{SAS} tool \emph{evselect} and background flares were filtered out prior to spectral extraction. The event files
were then checked for pile-up, using \emph{SAS} tool \emph{epatplot} and as EPIC-pn has a fast readout time of 5.7 milliseconds for
small window mode it was found that there is no problem of pile up in the source spectrum. 

For the extraction of the source spectra and lightcurves, circular region of 40 arcsecond radius was selected around the
source centroid. For the background spectra and light curve extraction, two off-source regions were used with a radius of 40 
arcseconds. Spectral extraction was done with PATTERN $\leq$ 4 and FLAG=0. For reprocessing of the data and generation of the response 
files updated version of the current calibration files (ccf) \footnote[1]{http://xmm2.esac.esa.int/external/xmm\texttt{\_sw\_cal}/calib/ccf.shtml} 
was used.

\section{Pulse Profiles}
The energy dependent pulse profiles from this observation have already been reported in 
\citet{Krauss07}. As a prerequisite to the phase resolved spectrsocopy reported in the 
next section, we have repeated the period analysis. We have derived a pulse period
of 7.67548 seconds which is consistent with \citet{Krauss07}. However, the energy
resolved pulse profiles shown in the left panel of Figure~\ref{lineflux} in 
different energy bands show substantially more fine structures. \\

Pulse profile of 4U~1626--67 is known to have strong energy dependence \citep{Rappaport77, Joss78, Pravdo79,
Elsner80, Kii86, Levine88, Mihara95, Angelini95, Krauss07, Jain08, Iwakiri12}.
\citet{Angelini95,Krauss07} and \citet{Jain08,Jain10} also reported
pulse profiles similar to Figure~\ref{lineflux} during the spin-down phase of the 
pulsar. These profiles were significantly different from that obtained in two spin-up
eras \citep{Beri14}.

\section{SPECTROSCOPY}

\subsection{Pulse Phase averaged Spectroscopy}

Pulse-phase averaged spectral analysis was performed using the mean spectrum extracted from the EPIC-pn data,
which was then grouped using ftool \emph{grppha} with minimum 500 counts/bin. The spectral fitting was
done using \emph{Xspec Version}: 12.8.0 \citep{Arnaud96}. The energy band chosen for fit was 0.3-10.0~keV. \\
 The continuum of the phase averaged spectrum is well described by a soft thermal component and a power law 
\citep{Pravdo79,Kii86,Angelini95,Owens97,Orlandini98,Schulz01,Krauss07,Jain10,Camero12,Iwakiri12}. The two 
components cross over at 0.8~keV. 
We used \emph{bbodyrad} model for fitting the soft part of the spectrum.

\begin{figure*}
\includegraphics[height=5.5in, width=5.5in, angle=-90]{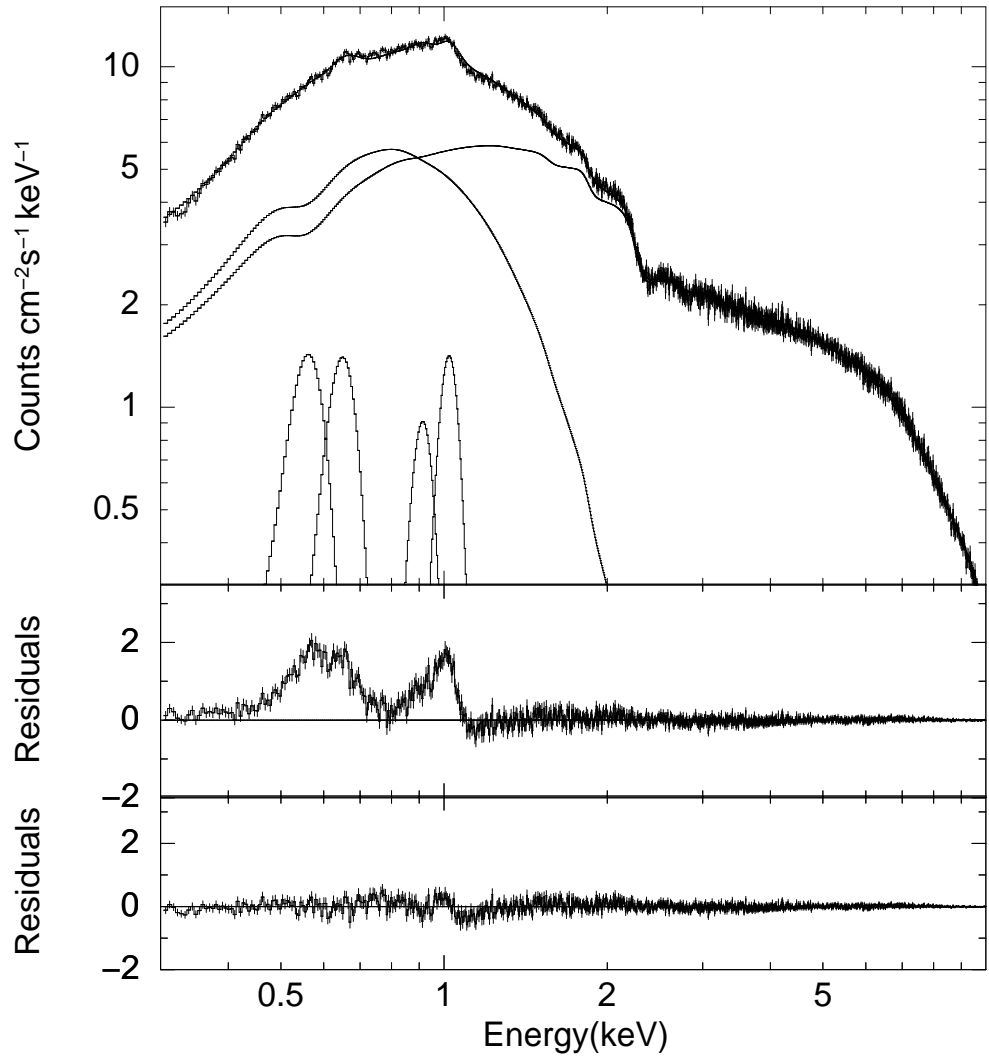}
 \caption{Data and folded model of 4U~1626--67 from \emph{XMM-Newton}~EPIC-pn. Upper panel shows the phase averaged spectrum. 
     The thick line is the best fit model obtained by fitting the continuum (consisting bbody and power-law). 
     Middle panel show the residuals ($Counts~cm^{-2}s^{-1}keV^{-1}$) obtained, using only continuum models and the lower panel are the residuals
     ($Counts~cm^{-2}s^{-1}keV^{-1}$) after fitting gaussians at 0.568~keV, 0.653~keV, 0.915~keV and 1.02~keV energies along with continuum models. }
\label{Average}
\end{figure*}

Using only the continuum model shows significant excess in the form of emission line features around 0.5~keV and 1.0~keV 
(see Figure~\ref{Average}). The \emph{RGS} data from the same \emph{XMM-Newton} observation was used by \citet{Krauss07}, 
where they found four emission lines at 0.568, 0.653, 0.915, 1.02~keV. All the four
lines were found to be narrow ($\sigma$ $\sim$~1eV).
Following those results we added four Gaussian lines with energies fixed at 0.568~(O~VII $He_{\alpha}$), 0.653~(O~VIII $Ly_{\alpha}$),
 0.915~(Ne~IX $He_{\alpha}$) and 1.02~keV~(Ne~X $Ly_{\alpha}$) to the continuum model,
 with their line widths fixed to $\sigma$=1~eV. On adding these lines an improved good fit was obtained with 
reduced $\chi^{2}$~($\chi^{2}_{\nu}$) of 1.175 for 1224 dof. The best fit parameters are given in Table-1. These fit parameters
are in good agreement with the values reported by \citet{Krauss07} using \emph{RGS} data from the same observation.

         \begin{table}
      	 \caption{Best-fitting parameters of Phase averaged spectrum of 4U1626-67.}
         \begin{tabular}{ l  l }
         \hline
         \hline
           
         Parameter & Model Values  \\ 
         N$_H$ (10${^2}{^2}$atoms cm$^{-2}$) & $0.102 \pm{0.001}$  \\
    	 
         PowIndex ($\Gamma$)  & $0.793\pm{0.003}$ \\ 
         
        $N_{PL}^a$  & $0.00656\pm{.00004}$  \\

         BBody~(kT)~keV  & $0.229\pm{0.001}$ \\
        
         $N_{BB}^b$  &  $441\pm{18}$  \\
         $\rm{Line Flux}^c$    \\        
         $\rm{O~VII~(0.568~keV)}$  & $6.4\pm{0.1}$ \\
       
         $\rm{O~VIII~(0.653~keV)}$  & $3.7\pm{0.1}$ \\
        
         $\rm{Ne~IX~(0.915~keV)}$ &  $1.36\pm{0.09}$ \\
        
         $\rm{Ne~X~(1.02~keV)}$  & $1.88\pm{0.07}$ \\
  \hline
  \end{tabular}

 \bigskip

{\bf{Notes}}: Errors quoted are for the 68 $\%$ confidence range. \\
              a $\rightarrow$ Powerlaw normalisation~($N_{PL}$) is in units of $\rm{photons~cm^{-2}~s^{-1}~keV^{-1}}$ at 1~keV \\
              b $\rightarrow$ Blackbody normalisation~($N_{BB}$) is in units of $\rm{photons~cm^{-2}~s^{-1}~keV^{-1}}$ \\
              c $\rightarrow$ Gaussian normalisation is in units of $10^{-4}$$\rm{photons~cm^{-2}~s^{-1}}$ \\

  \end{table}

\subsection{Phase-Resolved Spectroscopy}

\begin{figure*}
\centering
\begin{minipage}{0.45\textwidth}
\includegraphics[height=6.5in,width=8.5cm]{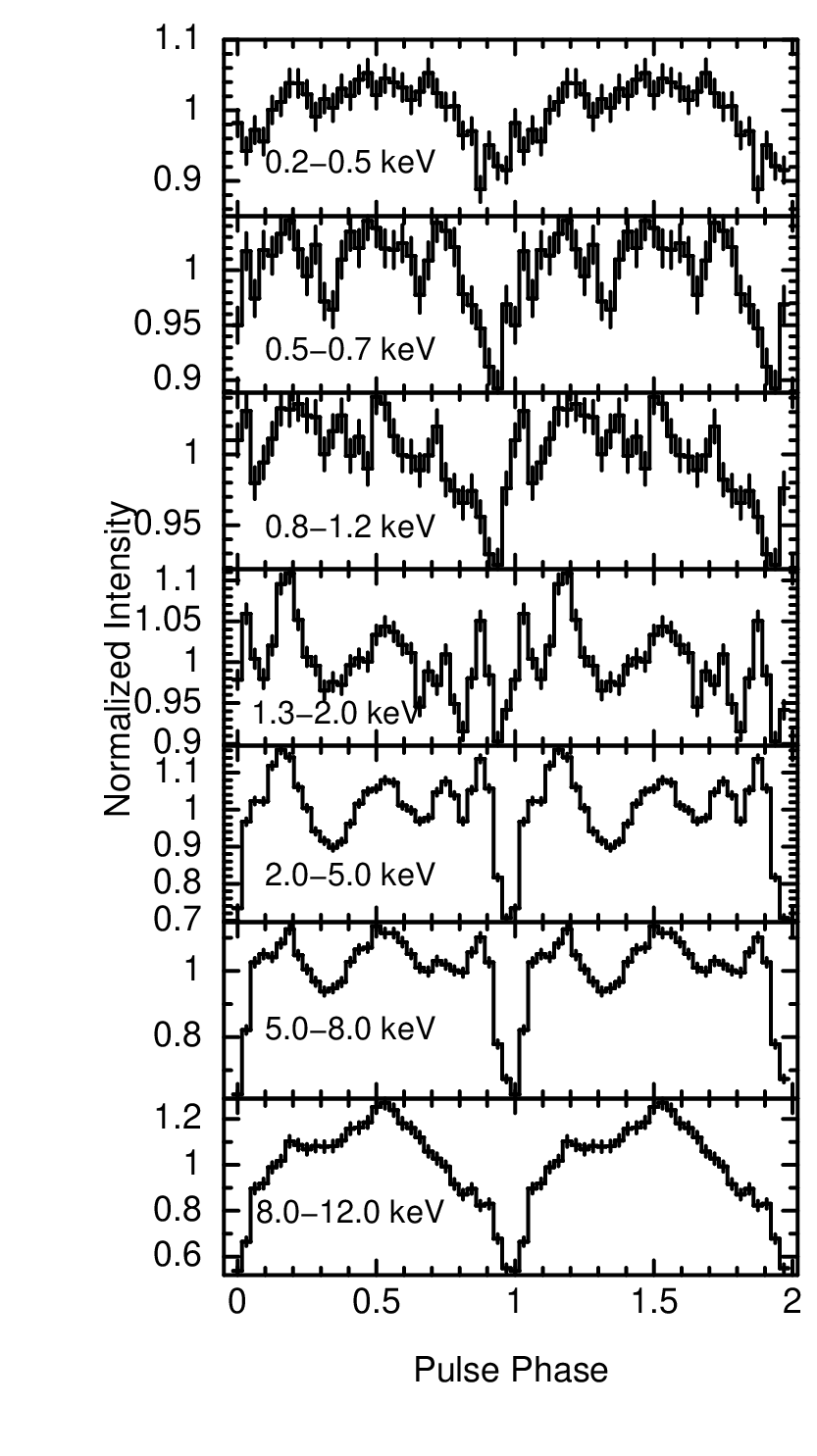}
\end{minipage}
\hspace{0.05\linewidth}
\begin{minipage}{0.45\textwidth}
\includegraphics[height=6.5in,width=8.5cm]{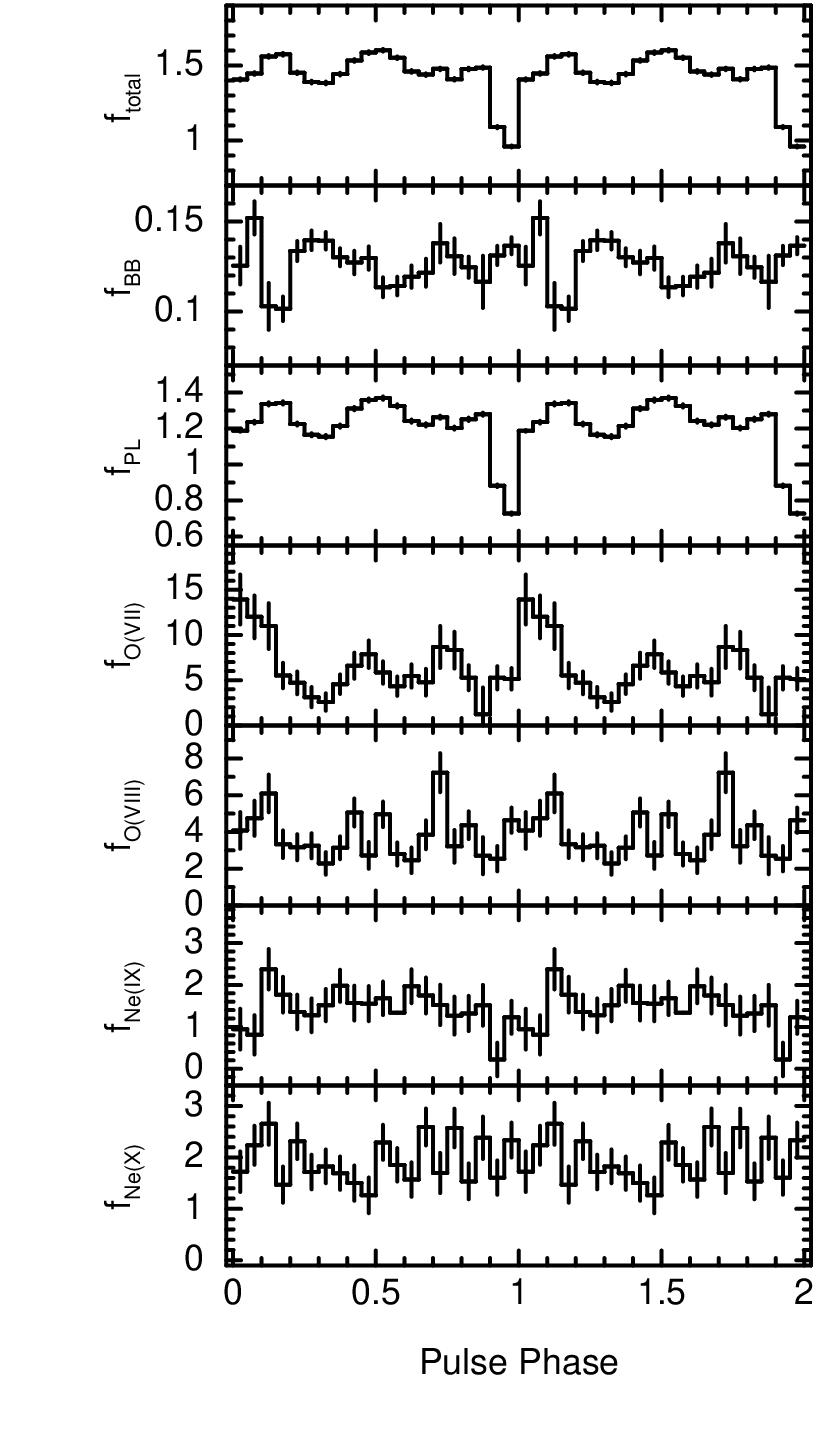}
\end{minipage}
\caption{The left-hand panels show the energy resolved pulse profiles created using EPIC-pn data and binned into 32 phasebins. The
right panel shows the variation of flux~(f) of continuum and the low energy emission lines near 1~keV with the pulse phase.
The units of all continuum fluxes~($f_{total}$, $f_{BB}$, $f_{PL}$)~are in $10^{-10}$$ergs~cm^{-2}~sec^{-1}$, 
while the line fluxes are measured in units of
$10^{-4}$$photons~cm^{-2}~sec^{-1}$. All the errors were estimated with 1$\sigma$ confidence.}
\label{lineflux}
\end{figure*}

The energy dependence of the pulse profiles in Figure~\ref{lineflux}, suggests a clear dependence of the energy spectrum
on the spin phase. A strong phase dependence of the continuum of 4U~1626--67 has been reported by \cite{Pravdo79}, \cite{Kii86},
\citet{Coburn00} and \citet{Iwakiri12} using the phase resolved spectra acquired by \emph{HEAO-1}, \emph{Tenma}, \emph{RXTE} 
and \emph{Suzaku} respectively.

In order to investigate the correlation between the features observed in the pulse profiles with Neon and Oxygen  
emission lines in the spectrum of 4U~1626--67 we performed a pulse-phase resolved spectroscopy. For the phase resolved spectral
analysis, the spectra were extracted with a phase bin of 0.05, using the 'phase' filter of \emph{ftools} \footnote{http://heasarc.gsfc.nasa.gov/ftools/}
task \textsc{xselect}.
We used the same background spectrum and response files as were used for the phase-averaged spectrum. X-ray spectra in
different phases were grouped using \emph{ftools} task \textsc{grppha} with minimum 250 counts per bin. For fitting the phase-resolved spectra
we followed exactly the same technique as we opted for the phase-averaged spectrum. \\

\begin{figure*}
\centering
\includegraphics[height=6.0in,width=11.5cm]{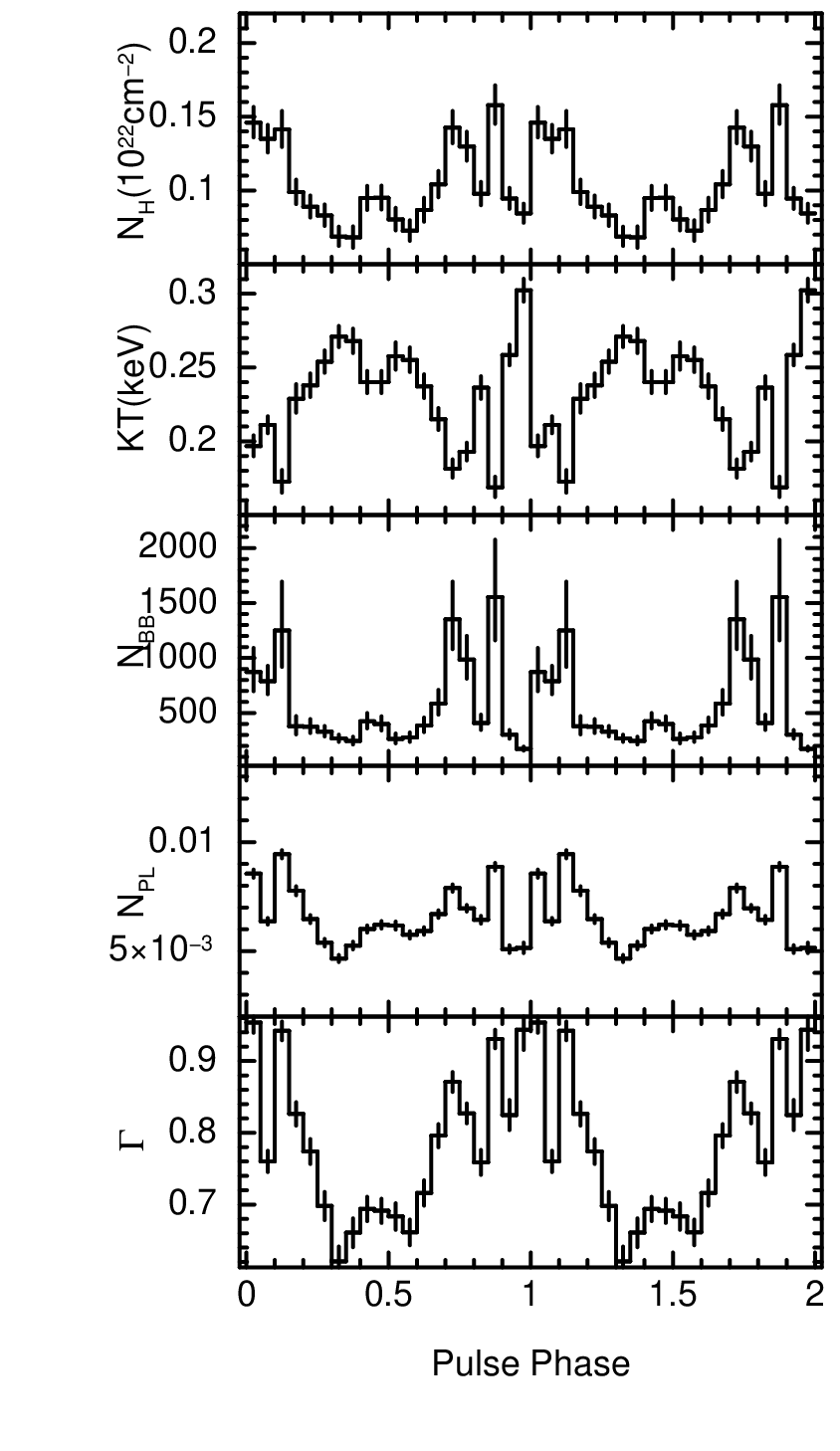}
\caption{Variation of spectral parameters of continuum accross the pulse phase for 4U~1626--67. 
The units of $N_H$ (equivalent hydrogen column density) is $10^{22}atoms~cm^{-2}$.
 Power-law ($N_{PL}$) and blackbody normalisations ($N_{BB}$) is in units of $photons~keV^{-1}~cm^{-2}~s^{-1}$.
 All the values are in 1$\sigma$ confidence.}
\label{continuum}
\end{figure*}

The continuum parameters (Figure~\ref{continuum}) show variation with the pulse phase.
The equivalent hydrogen coulmn density ($N_H$) showed significantly higher value compared
to the average value in the phase range 0.7-1.15 with some fluctuation. Temperature
of blackbody varied within the range of $\sim$ 0.16-0.30~keV showing lower value
around phase 0.7-1.15. The normalisation of blackbody also exhibits a behaviour that
is anticorrelated with its temperature profile.
However, we note that these spectral parameters can have degeneracy between them and the 
rapid fluctuations of these parameter as seen in Figure~\ref{continuum} does indicate the 
presence of significant degeneracy. We therefore do not comment further on this.
The powerlaw index however is determined from a broader energy range and shows lesser
bin to bin fluctuation. We find a pulse phase dependence of the power-law index with 
spectrum being hardest near the pulse peak and it being soft near the dip of the 
pulse profile. \\

Right panel in Figure~\ref{lineflux} shows the behaviour of total flux, blackbody flux, power law flux, Ne and O line intensities as 
function of the pulse phase. The total and power law flux show a sharp dip at around phase 1.0 along with two shallow dips
near phase 0.3 and 0.7. This trend is quite similar to that of its pulse profile.
 Blackbody flux also shows some structures in its profile with a dip between 0.2-0.3 phase.
 The behaviour of line fluxes shown in Figure~\ref{lineflux} is compared to the features in its pulse profile.\\

1.)The O~VII~(0.568~keV) emission line shows strong variability by more than a factor
of 4. This line is in maximum between the phase 0.0-0.15 and then decreases 
reaching a minimum intensity at around phase of 0.3. In the rest of the pulse phase also it shows some variation.
It is interesting to note that peak intensity of this line is coincident with dip in the pulse profile. \\
2.) The O~VIII~(0.653~keV) shows some variation with spin phase but without any clear structures in it. \\
3.) Ne~IX~(0.915~keV) shows some variation with spin phase with a possible correlation with the pulse 
profile in 0.8-1.2~keV energy band. \\
4.) The  Ne~X~(1.021 keV) line is consistent with no pulse phase dependence. \\

\begin{figure*}
\includegraphics[height=6.5in, width=4.5in]{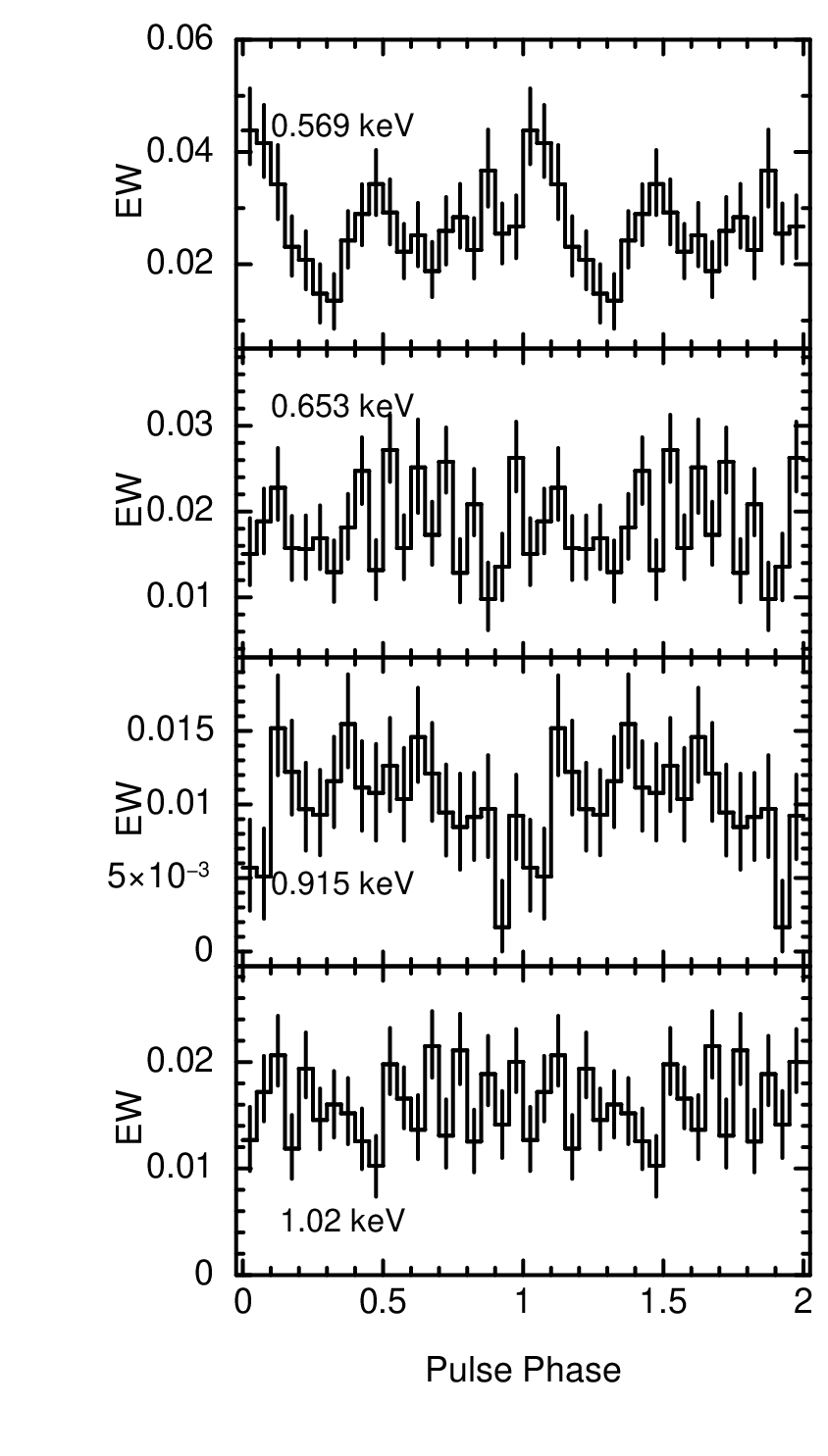}
\caption{Equivalent widths (EW) of Ne and O lines as the function of phase for 4U1626-67. EW is measured
in units of keV. The errors were estimated with 1$\sigma$ confidence.}
\label{eqwidth}
\end{figure*}

The equivalent width (EW) of these emission lines are shown in Figure~\ref{eqwidth}. The EW of
0.568~keV line also shows a maximum around phase 0.0-0.15 with a subsequent decreasing trend till phase 0.3. Some variation
in the EW of this line is seen throughout the pulse phase. The EW of 0.615~keV line also show variation with phase.
 A dip in the EW profile of 0.915~keV emission line is coincident with the main dip observed in the
pulse profile as well as its flux profile. However, the EW of 1.02~keV emission line is almost constant throughout its pulse phase. \\

\begin{figure*}
\includegraphics[height=4.5in, width=4.5in, angle=0]{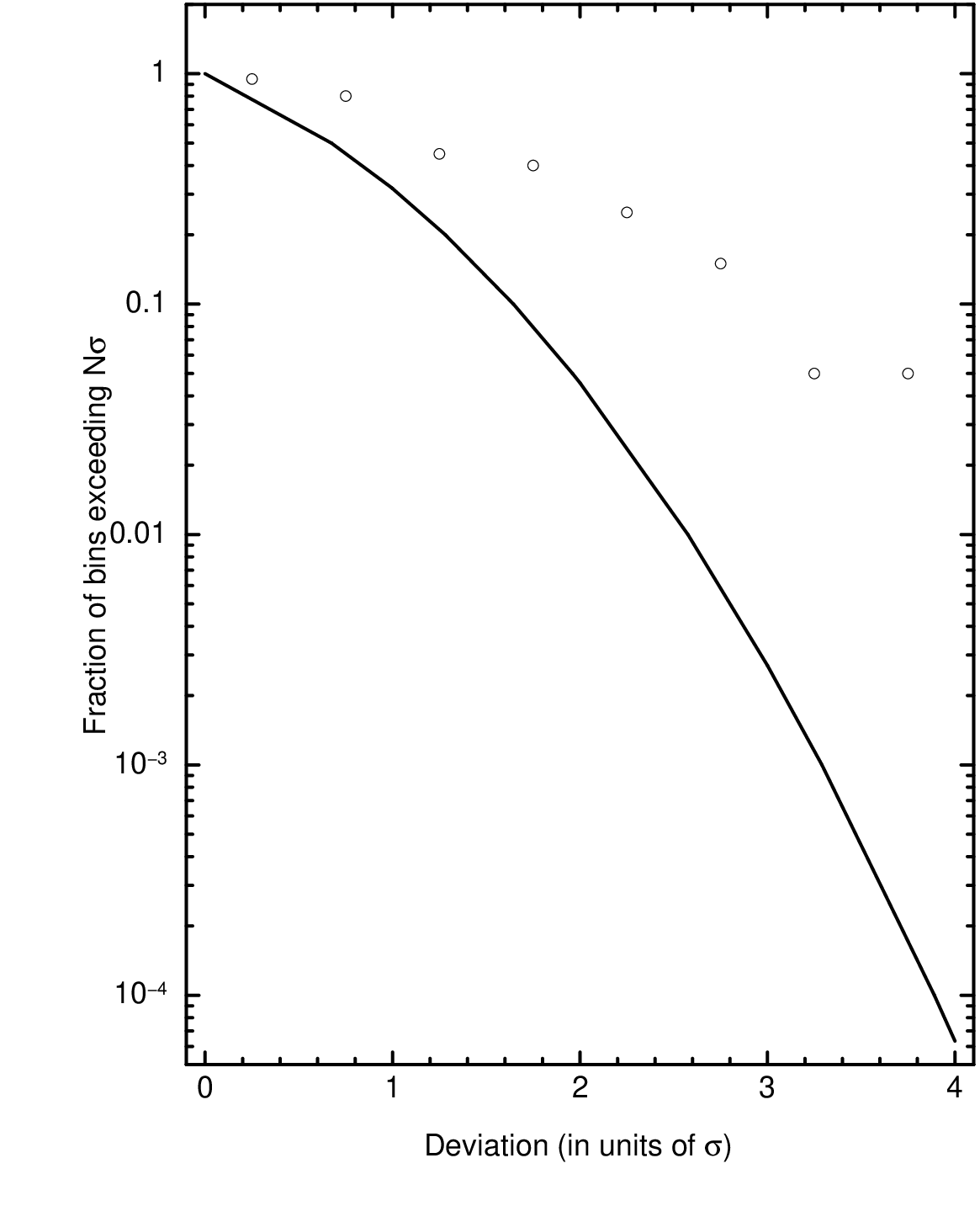}
\caption{The deviation of the flux of the O~VII line with pulse phase with respect to the average flux is
shown here. The fraction of the number of bins that deviates by more than N$\sigma$ is plotted as a function of N.
The same for purely statistical variation is shown with a line. The y-axis is in log scale.}
\label{histogram}
\end{figure*}

\begin{figure*}
\centering
\begin{minipage}{0.45\textwidth}
\includegraphics[height=5.5in, width=3.5in, angle=0]{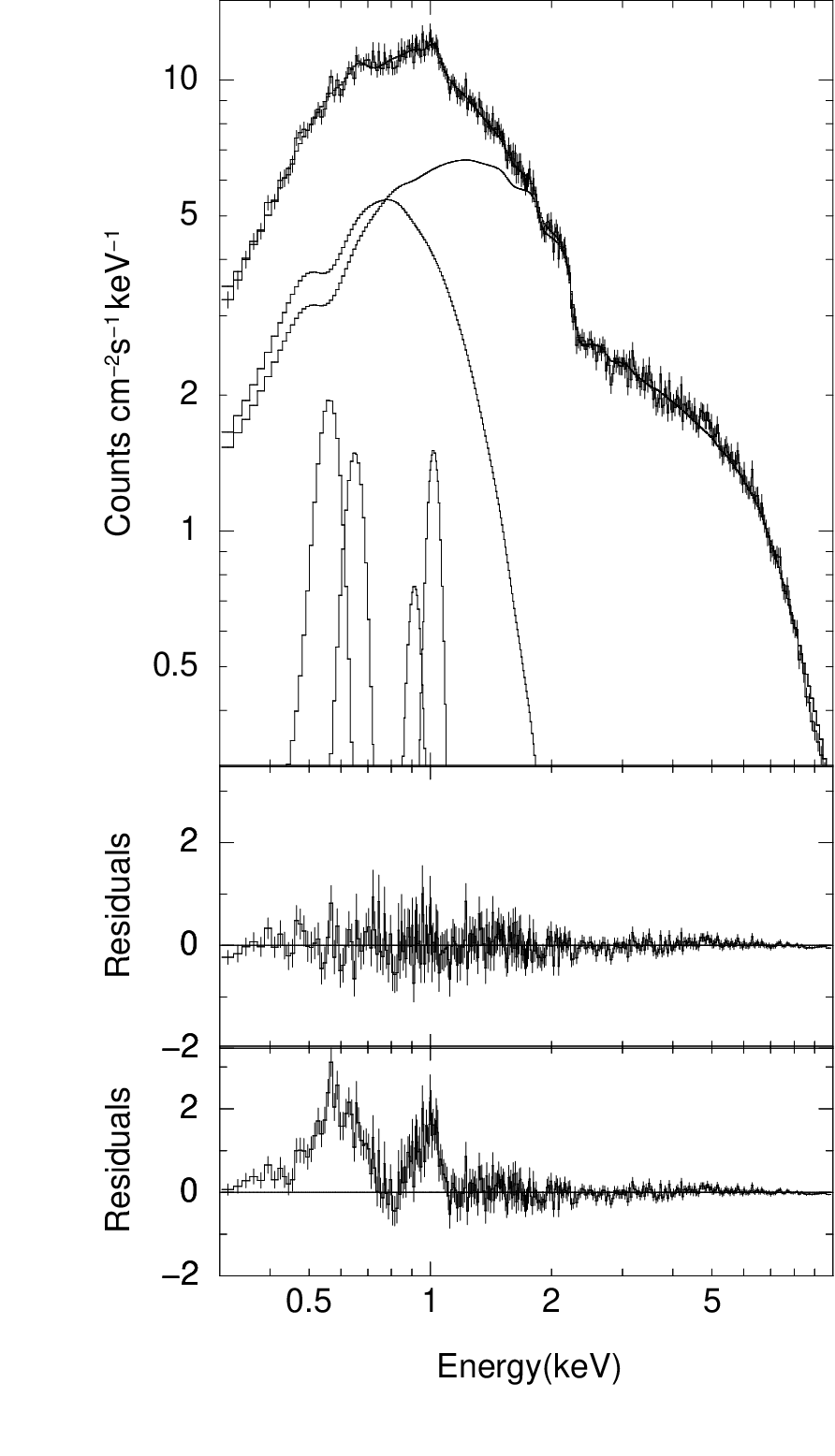}
\end{minipage}
\hspace{0.05\linewidth}
\begin{minipage}{0.45\textwidth}
\includegraphics[height=5.5in, width=3.5in, angle=0]{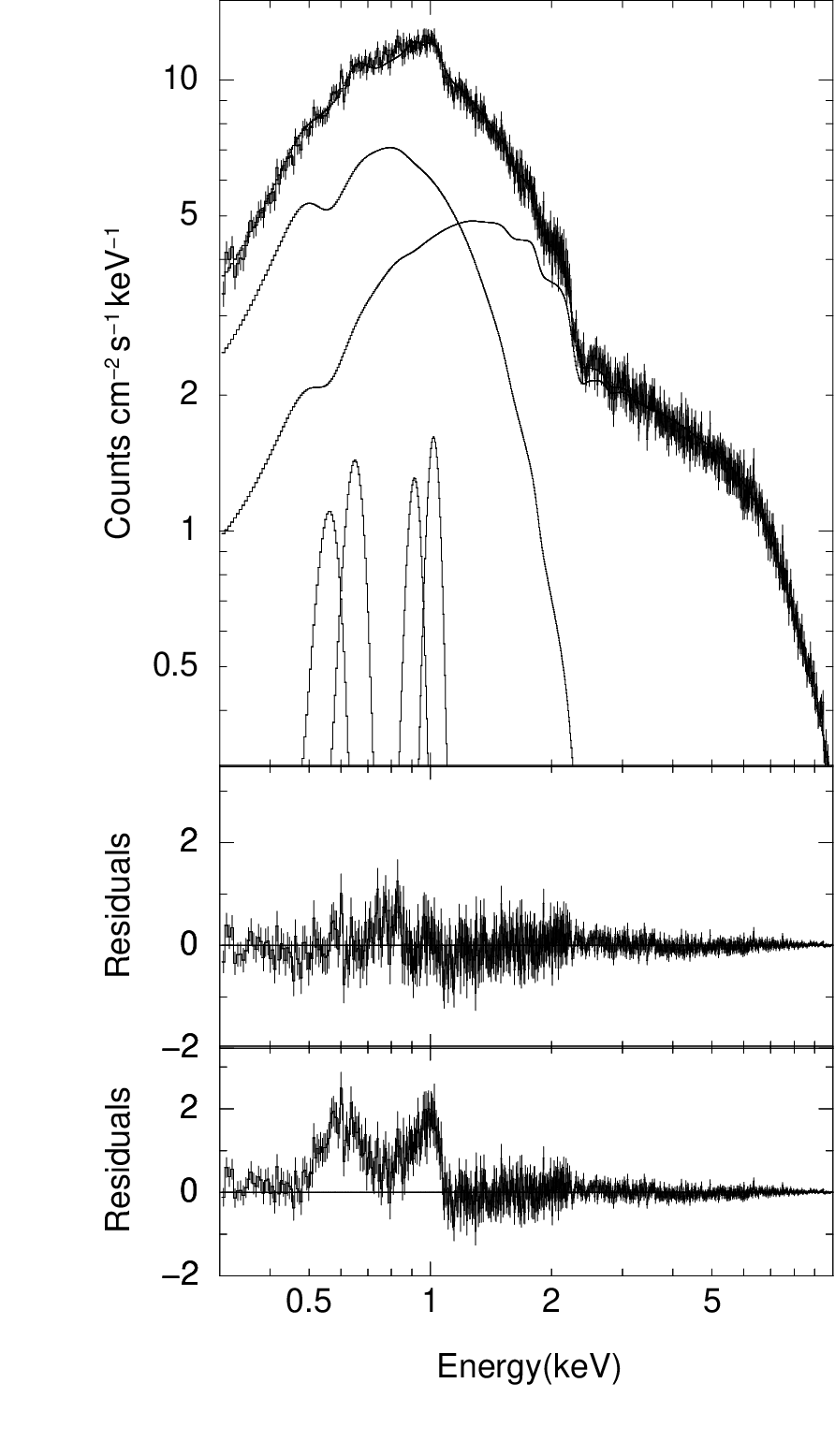}
\end{minipage}
\caption{In the left, panel-1 shows data and the best fit models of the spectrum extracted in (0.0-0.15) phase range, the middle panel 
shows the residuals ($Counts~cm^{-2}sec^{-1}keV^{-1}$) for the best fit and the bottom panel
 shows the residuals after setting line fluxes to zero. Similarly, in the right, the top panel shows data and the best fit models of the spectrum extracted 
 in (0.15-0.4) phase range, second panels contains the residuals obtained with best fit while the bottom panel has residuals 
with normalisations set to zero of line fluxes. Note the difference of the flux of the first emission line around 0.569~keV.}
\label{source0-0.15}
\end{figure*}

To evaluate the statistical significance of variation of the O~VII~(0.568 KeV) line with spin phase, 
we fixed the norm of the line 
in the phase resolved spectra with the value obtained from the phase-averaged spectrum. 
Fixing the line intensity to its phase averaged
 value resulted into an increase in the chi-squared ($\chi^{2}$) by upto 36 for 230 degrees of freedom (dof). In 
 5 of the 20 phase bins, the chance probability for the increase in $\chi^{2}$ is determined
by F-test to be less than $10^{-4}$ and in one of the bins it is as low as $5\times10^{-8}$,
which clearly demonstrates the variation of the O~VII emission line.  \\
A constant fitted to the flux (Figure~\ref{lineflux}) and equivalent width (Figure~\ref{eqwidth}) of the O~VII line
returned $\chi^{2}$ values of 46 and 32 respectively for the 20 phase bins, which is another measure
of the significance of the flux variability. \\
To further inspect the significance of the line intensity variation, we have plotted
the fraction of the number of phase bins for which the line flux differs from the
average value by more than N${\sigma}$ as a function of N. The same is compared
with what is expected without any intrinsic variability of the line (Figure~\ref{histogram}).
For any phase bin there is only 0.27~$\%$ chance probability for a 3~$\sigma$ deviation
while in three of the 20 phase bins in Figure~\ref{histogram} the line flux deviates from the pulse
averaged value by more than 3 $\sigma$ showing that the variability is statistically significant. \\
% For any phase bin there is only 0.27~$\%$ chance probability
% that the measured flux will exceed 3~$\sigma$ limit, however, from Figure~\ref{histogram}
% it is clear that three out of 20 phase bins~(15~$\%$)
% exceed 3~$\sigma$ limit
% This strongly implies that variability is statistically significant.   \\
We have also made a direct comparison of the spectra extracted in the pulse phase ranges
0.0-0.15 and 0.15-0.4. At these phases, flux of this line shows maximum variation. 
In Figure~\ref{source0-0.15}, we have shown the best fitted spectra along with the model components in top
panels for these two phase ranges. In the middle panels of the respective figures, we have
shown the residuals obtained for the best fit and in the bottom panels the residuals are shown
after setting the model line fluxes to zero. The residuals in the bottom panels clearly show
the flux difference for the 0.568 keV line.

To further investigate significance of the detected variation of the O~VII
line flux we simulated 10,000 spectra using the model parameters of the phase
averaged spectrum, with an exposure time typical of a phase resolved spectra
(2921~seconds). In each simulation, all the spectral parameters were taken
to be same as their phase averaged values. Thereafter, these simulated spectra
were fitted using the same model allowing all but the energy and widths of the
four lines to vary. A constant fitted to the flux of O~VII line returned a $\chi^2$ value of
8586 for 10,000 data points. We found that standard deviation of the O~VII line flux
variation in the simulated spectra was merely 15.2$\%$, and 90$\%$ of the simulation results
had flux within -36$\%$ and +40$\%$ of the average value. Thus the variation of line flux found in the
simulated spectra is quite low compared to upto a factor of 4 variation observed in
the O~VII line flux, indicating that degeneracy between the spectral parameters
did not result into values of the narrow line fluxes.  \\

\begin{figure*}
\includegraphics[height=5.5in, width=3.5in, angle=0]{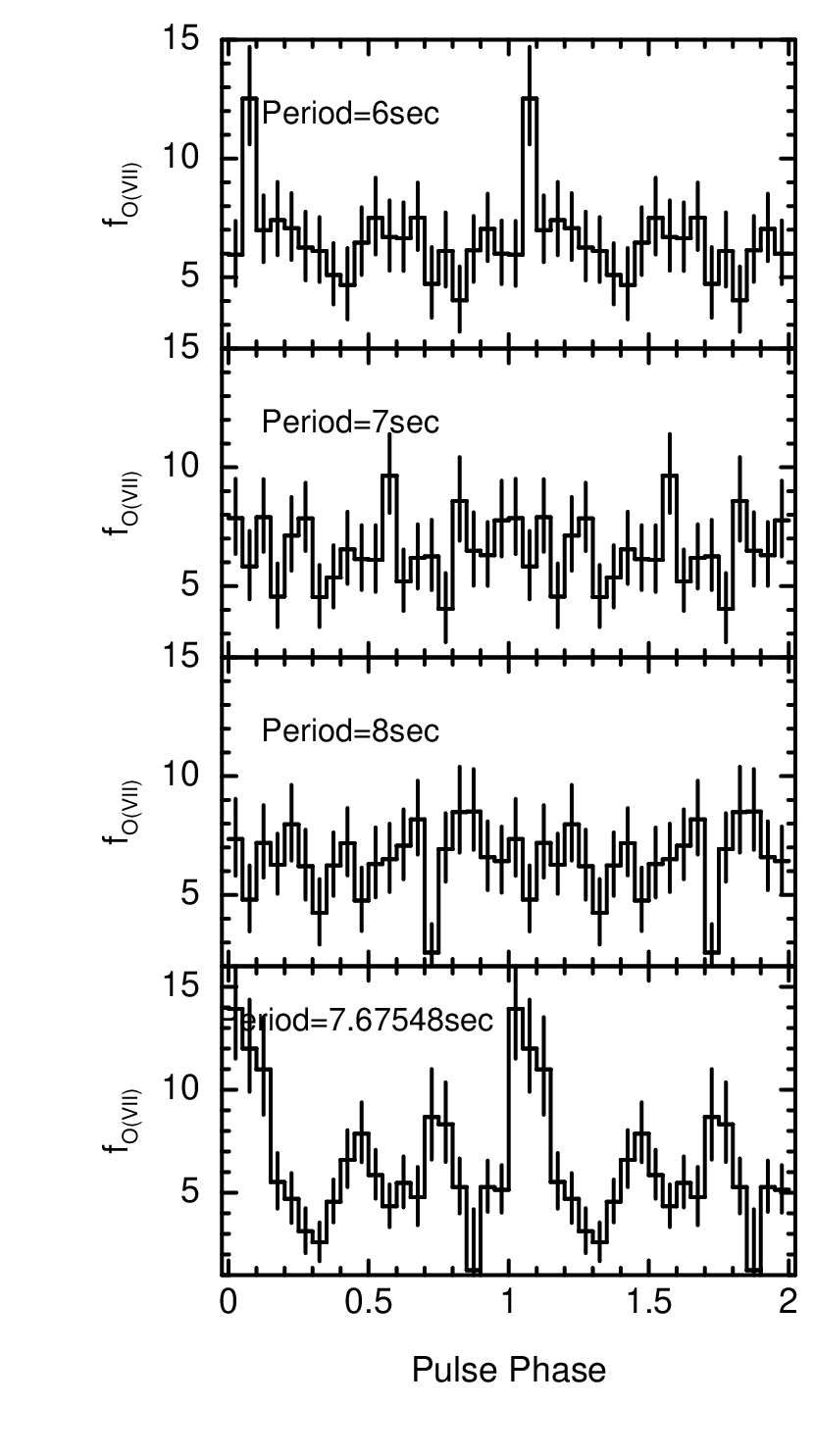}
\caption{\bf{{The variation of flux~(f) of O~VII emission line accross the pulse phase 
using different trial periods in comparison to the
true period=7.67548~seconds. 
The line flux is measured in units of
$10^{-4}$$photons~cm^{-2}~sec^{-1}$. All the errors were estimated with 1$\sigma$ confidence.}}}
\label{trial}
\end{figure*}

Moreover, we have also performed similar analysis with different trial pulse periods of 6,~7 and 8~seconds.
The phase resolved spectra created using these trial periods were fitted using the same model.
We did not find any significant variation of O~VII line flux at these arbitrary trial periods.
The constant fitted to the flux of O~VII line flux shown in Figure~\ref{trial} returned $\chi^{2}$ values of
16,~16 and 20 for 19 dof with the periods 6,~7 and 8 seconds respectively, compared to 46 for \emph{XMM-Newton} data.
This probe further strengthens our result of variation of O~VII line flux accross the 
pulse phase. \\

We also created pulse profiles in very narrow energy bands (of range of about 0.1~keV) 
to compare the shape of the profiles near 0.568~keV with that of the higher and lower energy continuum.
However, because of the fact that the equivalent width~(EW) of these emission lines is quite low (10-30~eV) in comparison
to the energy resolution of about 100~eV and
  these line energies contribute to a small fraction
of emission even for a band width of 100~eV, the difference between the pulse profiles was not very prominent, though
we could observe some differences in the sharpness of the dips in the profiles. 
This was also consistent with our results obtained from pulse phase resolved spectroscopy.

\section{DISCUSSIONS AND CONCLUSIONS}

In the present work, we have performed a detailed pulse-phase resolved spectroscopy of the unique high magnetic 
field LMXB pulsar 4U~1626--67 using data from EPIC-pn onboard \emph{XMM-Newton} observatory to investigate the flux variation
of complex emission lines arising from the highly ionised species of Ne and O \citep{Angelini95,Owens97,Krauss07}
in order to investigate the origin of this unusual complex of emission lines.
\emph{Chandra}-HETG observations of 4U 1626-67 made in 2001 revealed a double peaked nature of these lines \citep{Schulz01}
interpreted as Doppler pairs of broadened lines of hydrogenic and He like Ne and O.
The Doppler velocity corresponding to the widths of these lines is of the same order as Keplarian velocities in the
accretion disks. \citet{Schulz01} have also mentioned the less likely possibility of Doppler shifted lines originating 
in a bipolar disk outflow. In the case of an accretion disk origin, lines of different elements and ionisation
states will be produced in different annular regions of the accretion disk, at the corresponding values of the
ionisation parameter. 
\\

Emission lines, especially the iron fluorescence line at 6.4~keV is ubiquituous among the X-ray pulsars and has various 
origins like the acretion disk, strong stellar wind of the high mass companion, accretion stream from the companion star, 
outflowing disk of the Be-star companions etc. The line flux is usually found to be variable, both with the continuum
flux and with the absorption column density \citep{Inoue85, Makishima86, Kotani99, Naik03, Furst11}. However, in the accretion powered pulsars,
the line flux is often found to have no pulse phase dependence \citep{Paul02, Lei09, Tony11, Suchy12}. An exception is a strong pulse phase
dependence of the 6.4~keV line seen in Hercules~X-1 \citep{Vasco13} in which the line flux is found to be minimum 
near the pulse peak and there is a remarkable similarity between the pulse phase dependence of the iron line flux and
the power-law photon index.
\\

Pulse phase resolved spectroscopy of 4U~1626--67 carried out in 20 bins show
a strong flux variation of O~VII~(0.568~keV) line with pulse phase. We have also shown
a comparsion of spectrum in two phase ranges (0.0-0.15) and (0.15-0.4) where the
flux of this line shows maximum difference. However, we would like to add a
caveat that if a part of the line of sight absorption column is local to the source
with a pulse phase dependent ionisation state of the material, the same can introduce
some unforseen artifacts in the phase resolved measurements of the line parameters.
Unfortunately, it is not possible to investigate this with a pulse phase resolved
spectroscopy of the \emph{RGS} data.~The two lines O~VIII~(0.653~keV) and Ne~IX~(0.915~keV) 
show weak variation with phase while the Ne~X~(1.021~keV) line is consistent with being non-varying.
\\

In 4U~1626--67, we found the O~VII line flux to vary by factor of about four (maximum/minimum; Figure~\ref{lineflux}: $4^{th}$ panel
on the right column), significantly larger compared to the relative variation of total flux ( Figure~\ref{lineflux}: top panel
on right column). This indicates that the line flux variation is unlikely to be due to varying illumination of flat accretion disk.
The dimension of the accretion disk in 4U~1626--67, which has an orbital period of about 40~minutes is about 0.5~lt-sec which is 
also smaller than the variability time scale observed for the line flux, about 2 seconds. Thus, one possibility of the 
observed line flux variation could be warps in the accretion disk.
Milli-Hertz QPOs in the UV and optical lightcurves of 4U~1626--67 have been interpreted as signature
of warps in its accretion disk \citep{Chakrabarty01}. In the presence of warps, the visibility of the X-ray illuminated
part of the accretion disk can have significant variation with pulse phase. 
If any of the emission lines is produced at the same radius as the warp, it may result into a pulse phase dependent line flux. A 
very low upper limit of mass function obtained from pulse arrival 
time delay \citep{Levine88, Jain08} indicates a very small mass of the companion star or a
large angle of the orbital plane, or a combination of both. 
While much of the accretion disk is likely to be in the orbital plane, the line production region (inner accretion disk)
probably has warped structure and has a different inclination, which is also borne out from the mHz QPOs \citep{Chakrabarty01}. \\

Though the results presented here are indicative of an accretion disk with a warp around
a distance at which the O~VII emission line is produced, it is also possible to produce a large flux variation
of one emission line if the accretion disk sees a very different pulse profile compared to the observer 
and if the lines produced have different escape probability.
Since, only O~VII emission line showed strong variation accross the pulse phase, it may be possible
one of the component of triplet O~VII emission line may have higher escape probability. \\

The X-ray timing characteristics, like the pulse profile, the QPO's etc of 4U~1626--67 are known to be different in the spin-down
phase compared to spin-up phase \citep{Jain10, Beri14} which may also be caused by a difference in the inner accretion flow 
from a warped accretion disk in the spin-down phase. 
If this scenario is true, the Doppler characteristics of the O and Ne lines are expected to be different in the current spin-up phase of 4U~1626--67.
We also note that Her X-1 is known to have a warped accretion disk, 
but the pulse phase dependence of the iron line flux in Her X-1 is almost 
identical in the different superorbital phases \citep{Vasco13} which is caused by precession of the warped structure.
Thus the variation of the line flux in the two sources 4U~1626--67 and Her X-1 have different origin 
and the latter is explained to be due to reprocessing of continuum X-ray in the accretion coulmn with hollow cone geometry.

\section{Summary}

We have found the O~VII emission line in the spectrum of the accretion
powered pulsar 4U~1626--67 to vary with pulse phase, the only pulsar
after Her~X--1 to show a pulse dependent emission line flux.
The flux variation is by a factor of up to 4, much larger than the pulse fraction,
indicating the presence of warp in the accretion disk at the same radius at which
this line is produced by reprocessing of the X-rays emitted from the neutron star.
This is also the first X-ray spectroscopic evidence of an accretion disk warp in
an X-ray binary.

\section*{Acknowledgments}
A.B would like to thank all the members of the 2nd X-ray Astronomy workshop, held in IUCAA, Pune, India, 2013 and also
acknowledges IIT ropar for financial assistance and Raman Research Institute~(RRI) for providing local hospitality. 
The research has made use of 
data obtained from High Energy Astrophysics Science Archive Research Center (HEASARC), provided by NASA Goddard Space Flight Center.
We thank an anonymous referee for very constructive and 
useful comments, which improved the paper. 

%  \begin{thebibliography}{100}
\bibliography{bibtex}{}

\begin{thebibliography}{51}
\expandafter\ifx\csname natexlab\endcsname\relax\def\natexlab#1{#1}\fi

\bibitem[{{Angelini} {et~al}\mbox{.}(1995){Angelini}, {White}, {Nagase},
  {Kallman}, {Yoshida}, {Takeshima}, {Becker}, \& {Paerels}}]{Angelini95}
{Angelini} L., {White} N.~E., {Nagase} F., {Kallman} T.~R., {Yoshida} A.,
  {Takeshima} T., {Becker} C., {Paerels} F., 1995, \apjl, 449, L41

\bibitem[{{Arnaud}(1996)}]{Arnaud96}
{Arnaud} K.~A., 1996, in Astronomical Society of the Pacific Conference Series,
  Vol. 101, Astronomical Data Analysis Software and Systems V, {Jacoby} G.~H.,
  {Barnes} J., eds., p.~17

\bibitem[{{Beri} {et~al}\mbox{.}(2014){Beri}, {Jain}, {Paul}, \&
  {Raichur}}]{Beri14}
{Beri} A., {Jain} C., {Paul} B., {Raichur} H., 2014, \mnras, 439, 1940

\bibitem[{{Bildsten} {et~al}\mbox{.}(1994){Bildsten}, {Chakrabarty}, {Chiu},
  {Finger}, {Grunsfeld}, {Koh}, {Prince}, \& {Wilson}}]{Bildsten94}
{Bildsten} L., {Chakrabarty} D., {Chiu} J., {Finger} M.~H., {Grunsfeld} J.~M.,
  {Koh} T., {Prince} T.~A., {Wilson} R., 1994, in American Institute of Physics
  Conference Series, Vol. 304, American Institute of Physics Conference Series,
  {Fichtel} C.~E., {Gehrels} N., {Norris} J.~P., eds., pp. 294--298

\bibitem[{{Bildsten} {et~al}\mbox{.}(1997){Bildsten}, {Chakrabarty}, {Chiu},
  {Finger}, {Koh}, {Nelson}, {Prince}, {Rubin}, {Scott}, {Stollberg},
  {Vaughan}, {Wilson}, \& {Wilson}}]{Bildsten97}
{Bildsten} L. {et~al.}, 1997, \apjs, 113, 367

\bibitem[{{Camero-Arranz} {et~al}\mbox{.}(2010){Camero-Arranz}, {Finger},
  {Ikhsanov}, {Wilson-Hodge}, \& {Beklen}}]{Camero10}
{Camero-Arranz} A., {Finger} M.~H., {Ikhsanov} N.~R., {Wilson-Hodge} C.~A.,
  {Beklen} E., 2010, \apj, 708, 1500

\bibitem[{{Camero-Arranz} {et~al}\mbox{.}(2012){Camero-Arranz}, {Pottschmidt},
  {Finger}, {Ikhsanov}, {Wilson-Hodge}, \& {Marcu}}]{Camero12}
{Camero-Arranz} A., {Pottschmidt} K., {Finger} M.~H., {Ikhsanov} N.~R.,
  {Wilson-Hodge} C.~A., {Marcu} D.~M., 2012, \aap, 546, A40

\bibitem[{{Chakrabarty}(1998)}]{Chakrabarty98}
{Chakrabarty} D., 1998, \apj, 492, 342

\bibitem[{{Chakrabarty} {et~al}\mbox{.}(1997){Chakrabarty}, {Bildsten},
  {Grunsfeld}, {Koh}, {Prince}, {Vaughan}, {Finger}, {Scott}, \&
  {Wilson}}]{Chakrabarty97}
{Chakrabarty} D. {et~al.}, 1997, \apj, 474, 414

\bibitem[{{Chakrabarty} {et~al}\mbox{.}(2001){Chakrabarty}, {Homer}, {Charles},
  \& {O'Donoghue}}]{Chakrabarty01}
{Chakrabarty} D., {Homer} L., {Charles} P.~A., {O'Donoghue} D., 2001, \apj,
  562, 985

\bibitem[{{Coburn} {et~al}\mbox{.}(2000){Coburn}, {Heindl}, {Gruber},
  {Rothschild}, {Staubert}, {Wilms}, \& {Kreykenbohm}}]{Coburn00}
{Coburn} W., {Heindl} W.~A., {Gruber} D.~E., {Rothschild} R.~E., {Staubert} R.,
  {Wilms} J., {Kreykenbohm} I., 2000, in Bulletin of the American Astronomical
  Society, Vol.~32, AAS/High Energy Astrophysics Division \#5, p. 1213

\bibitem[{{Coburn} {et~al}\mbox{.}(2002){Coburn}, {Heindl}, {Rothschild},
  {Gruber}, {Kreykenbohm}, {Wilms}, {Kretschmar}, \& {Staubert}}]{Coburn02}
{Coburn} W., {Heindl} W.~A., {Rothschild} R.~E., {Gruber} D.~E., {Kreykenbohm}
  I., {Wilms} J., {Kretschmar} P., {Staubert} R., 2002, \apj, 580, 394

\bibitem[{{den Herder} {et~al}\mbox{.}(2001){den Herder}, {Brinkman}, {Kahn},
  {Branduardi-Raymont}, {Thomsen}, {Aarts}, {Audard}, {Bixler}, {den Boggende},
  {Cottam}, {Decker}, {Dubbeldam}, {Erd}, {Goulooze}, {G{\"u}del}, {Guttridge},
  {Hailey}, {Janabi}, {Kaastra}, {de Korte}, {van Leeuwen}, {Mauche},
  {McCalden}, {Mewe}, {Naber}, {Paerels}, {Peterson}, {Rasmussen}, {Rees},
  {Sakelliou}, {Sako}, {Spodek}, {Stern}, {Tamura}, {Tandy}, {de Vries},
  {Welch}, \& {Zehnder}}]{Herder01}
{den Herder} J.~W. {et~al.}, 2001, \aap, 365, L7

\bibitem[{{Elsner}, {Ghosh} \& {Lamb}(1980){Elsner}, {Ghosh}, \&
  {Lamb}}]{Elsner80}
{Elsner} R.~F., {Ghosh} P., {Lamb} F.~K., 1980, \apjl, 241, L155

\bibitem[{{F{\"u}rst} {et~al}\mbox{.}(2011){F{\"u}rst}, {Suchy}, {Kreykenbohm},
  {Barrag{\'a}n}, {Wilms}, {Pottschmidt}, {Caballero}, {Kretschmar},
  {Ferrigno}, \& {Rothschild}}]{Furst11}
{F{\"u}rst} F. {et~al.}, 2011, \aap, 535, A9

\bibitem[{{Giacconi} {et~al}\mbox{.}(1972){Giacconi}, {Murray}, {Gursky},
  {Kellogg}, {Schreier}, \& {Tananbaum}}]{Giacconi72}
{Giacconi} R., {Murray} S., {Gursky} H., {Kellogg} E., {Schreier} E.,
  {Tananbaum} H., 1972, \apj, 178, 281

\bibitem[{{Homer} {et~al}\mbox{.}(2002){Homer}, {Anderson}, {Wachter}, \&
  {Margon}}]{Homer02}
{Homer} L., {Anderson} S.~F., {Wachter} S., {Margon} B., 2002, \aj, 124, 3348

\bibitem[{{Inoue}(1985)}]{Inoue85}
{Inoue} H., 1985, \ssr, 40, 317

\bibitem[{{Iwakiri} {et~al}\mbox{.}(2012){Iwakiri}, {Terada}, {Mihara},
  {Angelini}, {Tashiro}, {Enoto}, {Yamada}, {Makishima}, {Nakajima}, \&
  {Yoshida}}]{Iwakiri12}
{Iwakiri} W.~B. {et~al.}, 2012, \apj, 751, 35

\bibitem[{{Jain}, {Paul} \& {Dutta}(2010){Jain}, {Paul}, \& {Dutta}}]{Jain10}
{Jain} C., {Paul} B., {Dutta} A., 2010, \mnras, 403, 920

\bibitem[{{Jain} {et~al}\mbox{.}(2008){Jain}, {Paul}, {Joshi}, {Dutta}, \&
  {Raichur}}]{Jain08}
{Jain} C., {Paul} B., {Joshi} K., {Dutta} A., {Raichur} H., 2008, Journal of
  Astrophysics and Astronomy, 28, 175

\bibitem[{{Joss}, {Avni} \& {Rappaport}(1978){Joss}, {Avni}, \&
  {Rappaport}}]{Joss78}
{Joss} P.~C., {Avni} Y., {Rappaport} S., 1978, \apj, 221, 645

\bibitem[{{Kii} {et~al}\mbox{.}(1986){Kii}, {Hayakawa}, {Nagase}, {Ikegami}, \&
  {Kawai}}]{Kii86}
{Kii} T., {Hayakawa} S., {Nagase} F., {Ikegami} T., {Kawai} N., 1986, \pasj,
  38, 751

\bibitem[{{Kotani} {et~al}\mbox{.}(1999){Kotani}, {Dotani}, {Nagase},
  {Greenhill}, {Pravdo}, \& {Angelini}}]{Kotani99}
{Kotani} T., {Dotani} T., {Nagase} F., {Greenhill} J.~G., {Pravdo} S.~H.,
  {Angelini} L., 1999, \apj, 510, 369

\bibitem[{{Krauss} {et~al}\mbox{.}(2007){Krauss}, {Schulz}, {Chakrabarty},
  {Juett}, \& {Cottam}}]{Krauss07}
{Krauss} M.~I., {Schulz} N.~S., {Chakrabarty} D., {Juett} A.~M., {Cottam} J.,
  2007, \apj, 660, 605

\bibitem[{{Lei} {et~al}\mbox{.}(2009){Lei}, {Chen}, {Qu}, {Song}, {Zhang},
  {Lu}, {Zhang}, \& {Li}}]{Lei09}
{Lei} Y.-J., {Chen} W., {Qu} J.-L., {Song} L.-M., {Zhang} S., {Lu} Y., {Zhang}
  H.-T., {Li} T.-P., 2009, \apj, 707, 1016

\bibitem[{{Levine} {et~al}\mbox{.}(1988){Levine}, {Ma}, {McClintock},
  {Rappaport}, {van der Klis}, \& {Verbunt}}]{Levine88}
{Levine} A., {Ma} C.~P., {McClintock} J., {Rappaport} S., {van der Klis} M.,
  {Verbunt} F., 1988, \apj, 327, 732

\bibitem[{{Makishima}(1986)}]{Makishima86}
{Makishima} K., 1986, in Lecture Notes in Physics, Berlin Springer Verlag, Vol.
  266, The Physics of Accretion onto Compact Objects, {Mason} K.~O., {Watson}
  M.~G., {White} N.~E., eds., p. 249

\bibitem[{{McClintock} {et~al}\mbox{.}(1977){McClintock}, {Bradt}, {Doxsey},
  {Jernigan}, {Canizares}, \& {Hiltner}}]{McClintock77}
{McClintock} J.~E., {Bradt} H.~V., {Doxsey} R.~E., {Jernigan} J.~G.,
  {Canizares} C.~R., {Hiltner} W.~A., 1977, \nat, 270, 320

\bibitem[{{McClintock} {et~al}\mbox{.}(1980){McClintock}, {Li}, {Canizares}, \&
  {Grindlay}}]{McClintock80}
{McClintock} J.~E., {Li} F.~K., {Canizares} C.~R., {Grindlay} J.~E., 1980,
  \apjl, 235, L81

\bibitem[{{Middleditch} {et~al}\mbox{.}(1981){Middleditch}, {Mason}, {Nelson},
  \& {White}}]{Middleditch81}
{Middleditch} J., {Mason} K.~O., {Nelson} J.~E., {White} N.~E., 1981, \apj,
  244, 1001

\bibitem[{{Mihara}(1995)}]{Mihara95}
{Mihara} T., 1995, PhD thesis, , Dept.~of Physics, Univ.~of Tokyo (M95), (1995)

\bibitem[{{Nagase}(1989)}]{Nagase89}
{Nagase} F., 1989, \pasj, 41, 1

\bibitem[{{Naik} \& {Paul}(2003)}]{Naik03}
{Naik} S., {Paul} B., 2003, \aap, 401, 265

\bibitem[{{Orlandini} {et~al}\mbox{.}(1998){Orlandini}, {Dal Fiume},
  {Frontera}, {Del Sordo}, {Piraino}, {Santangelo}, {Segreto}, {Oosterbroek},
  \& {Parmar}}]{Orlandini98}
{Orlandini} M. {et~al.}, 1998, \apjl, 500, L163

\bibitem[{{Owens}, {Oosterbroek} \& {Parmar}(1997){Owens}, {Oosterbroek}, \&
  {Parmar}}]{Owens97}
{Owens} A., {Oosterbroek} T., {Parmar} A.~N., 1997, \aap, 324, L9

\bibitem[{{Paczynski} \& {Sienkiewicz}(1981)}]{Paczynski81}
{Paczynski} B., {Sienkiewicz} R., 1981, \apjl, 248, L27

\bibitem[{{Paul} {et~al}\mbox{.}(2002){Paul}, {Dewangan}, {Sako}, {Kahn},
  {Paerels}, {Liedahl}, {Wojdowski}, \& {Nagase}}]{Paul02}
{Paul} B., {Dewangan} G.~C., {Sako} M., {Kahn} S.~M., {Paerels} F., {Liedahl}
  D., {Wojdowski} P., {Nagase} F., 2002, in 8th Asian-Pacific Regional Meeting,
  Volume II, {Ikeuchi} S., {Hearnshaw} J., {Hanawa} T., eds., pp. 355--356

\bibitem[{{Pravdo} {et~al}\mbox{.}(1979){Pravdo}, {White}, {Boldt}, {Holt},
  {Serlemitsos}, {Swank}, {Szymkowiak}, {Tuohy}, \& {Garmire}}]{Pravdo79}
{Pravdo} S.~H. {et~al.}, 1979, \apj, 231, 912

\bibitem[{{Prince} {et~al}\mbox{.}(1994){Prince}, {Bildsten}, {Chakrabarty},
  {kinsomnson}, \& {Finger}}]{Prince94}
{Prince} T.~A., {Bildsten} L., {Chakrabarty} D., {kinsomnson} R.~B., {Finger}
  M.~H., 1994, in American Institute of Physics Conference Series, Vol. 308,
  The Evolution of X-ray Binariese, {Holt} S., {Day} C.~S., eds., p. 235

\bibitem[{{Rappaport} {et~al}\mbox{.}(1977){Rappaport}, {Markert}, {Li},
  {Clark}, {Jernigan}, \& {McClintock}}]{Rappaport77}
{Rappaport} S., {Markert} T., {Li} F.~K., {Clark} G.~W., {Jernigan} J.~G.,
  {McClintock} J.~E., 1977, \apjl, 217, L29

\bibitem[{{Schulz} {et~al}\mbox{.}(2001){Schulz}, {Chakrabarty}, {Marshall},
  {Canizares}, {Lee}, \& {Houck}}]{Schulz01}
{Schulz} N.~S., {Chakrabarty} D., {Marshall} H.~L., {Canizares} C.~R., {Lee}
  J.~C., {Houck} J., 2001, \apj, 563, 941

\bibitem[{{Str{\"u}der} {et~al}\mbox{.}(2001){Str{\"u}der}, {Briel}, {Dennerl},
  {Hartmann}, {Kendziorra}, {Meidinger}, {Pfeffermann}, {Reppin}, {Aschenbach},
  {Bornemann}, {Br{\"a}uninger}, {Burkert}, {Elender}, {Freyberg}, {Haberl},
  {Hartner}, {Heuschmann}, {Hippmann}, {Kastelic}, {Kemmer}, {Kettenring},
  {Kink}, {Krause}, {M{\"u}ller}, {Oppitz}, {Pietsch}, {Popp}, {Predehl},
  {Read}, {Stephan}, {St{\"o}tter}, {Tr{\"u}mper}, {Holl}, {Kemmer}, {Soltau},
  {St{\"o}tter}, {Weber}, {Weichert}, {von Zanthier}, {Carathanassis}, {Lutz},
  {Richter}, {Solc}, {B{\"o}ttcher}, {Kuster}, {Staubert}, {Abbey}, {Holland},
  {Turner}, {Balasini}, {Bignami}, {La Palombara}, {Villa}, {Buttler},
  {Gianini}, {Lain{\'e}}, {Lumb}, \& {Dhez}}]{Struder01}
{Str{\"u}der} L. {et~al.}, 2001, \aap, 365, L18

\bibitem[{{Suchy} {et~al}\mbox{.}(2012){Suchy}, {F{\"u}rst}, {Pottschmidt},
  {Caballero}, {Kreykenbohm}, {Wilms}, {Markowitz}, \& {Rothschild}}]{Suchy12}
{Suchy} S., {F{\"u}rst} F., {Pottschmidt} K., {Caballero} I., {Kreykenbohm} I.,
  {Wilms} J., {Markowitz} A., {Rothschild} R.~E., 2012, \apj, 745, 124

\bibitem[{{Turner} {et~al}\mbox{.}(2001){Turner}, {Abbey}, {Arnaud},
  {Balasini}, {Barbera}, {Belsole}, {Bennie}, {Bernard}, {Bignami}, {Boer},
  {Briel}, {Butler}, {Cara}, {Chabaud}, {Cole}, {Collura}, {Conte}, {Cros},
  {Denby}, {Dhez}, {Di Coco}, {Dowson}, {Ferrando}, {Ghizzardi}, {Gianotti},
  {Goodall}, {Gretton}, {Griffiths}, {Hainaut}, {Hochedez}, {Holland},
  {Jourdain}, {Kendziorra}, {Lagostina}, {Laine}, {La Palombara}, {Lortholary},
  {Lumb}, {Marty}, {Molendi}, {Pigot}, {Poindron}, {Pounds}, {Reeves},
  {Reppin}, {Rothenflug}, {Salvetat}, {Sauvageot}, {Schmitt}, {Sembay},
  {Short}, {Spragg}, {Stephen}, {Str{\"u}der}, {Tiengo}, {Trifoglio},
  {Tr{\"u}mper}, {Vercellone}, {Vigroux}, {Villa}, {Ward}, {Whitehead}, \&
  {Zonca}}]{Turner01}
{Turner} M.~J.~L. {et~al.}, 2001, \aap, 365, L27

\bibitem[{{Vasco} {et~al}\mbox{.}(2013){Vasco}, {Staubert}, {Klochkov},
  {Santangelo}, {Shakura}, \& {Postnov}}]{Vasco13}
{Vasco} D., {Staubert} R., {Klochkov} D., {Santangelo} A., {Shakura} N.,
  {Postnov} K., 2013, \aap, 550, A111

\bibitem[{{Vaughan} \& {Kitamoto}(1997)}]{Vaughan97}
{Vaughan} B.~A., {Kitamoto} S., 1997, arXiv:astro-ph/9707105

\bibitem[{{Werner} {et~al}\mbox{.}(2006){Werner}, {Nagel}, {Rauch}, {Hammer},
  \& {Dreizler}}]{Werner06}
{Werner} K., {Nagel} T., {Rauch} T., {Hammer} N.~J., {Dreizler} S., 2006, \aap,
  450, 725

\bibitem[{{Wilkinson} {et~al}\mbox{.}(2011){Wilkinson}, {Patruno}, {Watts}, \&
  {Uttley}}]{Tony11}
{Wilkinson} T., {Patruno} A., {Watts} A., {Uttley} P., 2011, \mnras, 410, 1513

\bibitem[{{Wilson} {et~al}\mbox{.}(1993){Wilson}, {Fishman}, {Finger},
  {Pendleton}, {Prince}, \& {Chakrabarty}}]{Wilson93}
{Wilson} R.~B., {Fishman} G.~J., {Finger} M.~H., {Pendleton} G.~N., {Prince}
  T.~A., {Chakrabarty} D., 1993, in American Institute of Physics Conference
  Series, Vol. 280, American Institute of Physics Conference Series,
  {Friedlander} M., {Gehrels} N., {Macomb} D.~J., eds., pp. 291--302

\bibitem[{{Yi} \& {Vishniac}(1999)}]{Yi99}
{Yi} I., {Vishniac} E.~T., 1999, \apjl, 516, L87

\end{thebibliography}
\bibliographystyle{mn2e}

% 
% \begin{figure*}
% \includegraphics[height=5.5in, width=5.5in, angle=0]{histo.eps}
% \caption{Histogram of deviation from phase averaged value of flux measured for O~(VII) emission line at different pulse phases. 
% The y-axis shows the number of spectra that shows the variation within the specified range of bin while
% the x-axis refers to the measured deviation.
% For details please see text.}
% \label{histogram}
% \end{figure*}

\label{lastpage}
\end{document}